\documentclass[11pt, conference]{IEEEtran}
\usepackage{float}
\usepackage[dvipsnames]{xcolor}
\usepackage{tikz}
\usepackage{pgfplots}
\usetikzlibrary{patterns}
\usepackage{mathtools}
\usepackage{pst-node}
\usepackage{amsmath}

\date{August 1, 2016}

\def\layersep{2.0cm}
\usepackage{cite}

\usepackage{url}
\usepackage{comment}

\usepackage[dvipsnames]{xcolor}

\usepackage[
  pagebackref,
  bookmarksopen,
  bookmarksnumbered,
]{hyperref}
\hypersetup{
	pdfauthor={Toghiani-Rizi, Babak and Windmark, Marcus},
	colorlinks,
    linkcolor={red!70!black},
    citecolor={cyan!60!black},
    urlcolor={blue!80!black}
}

\usepackage[nameinlink]{cleveref}
\usepackage{bookmark}

\def\figureWidth{0.4}
\def\confusionMatrixWidth{0.68}

\pgfplotsset{compat=1.12}

\makeatletter
\def\ps@headings{%
  \def\@oddfoot{\scriptsize \@date\hfil \thepage}
  \def\@evenfoot{\scriptsize \@date\hfil \thepage}
}

\def\ps@IEEEtitlepagestyle{
  \def\@oddfoot{\scriptsize \@date\hfil \thepage}
  \def\@evenfoot{\scriptsize \@date\hfil \thepage}
}
\makeatother

\begin{document}
%

\title{Musical Instrument Recognition Using Their Distinctive  Characteristics \\ \huge{in Artificial Neural Networks}}

\author{
	\IEEEauthorblockN{Babak Toghiani-Rizi}
      \IEEEauthorblockA{
		bato9963@student.uu.se \\
        Department of Information Technology\\
        Uppsala University
      }
  \and
  \IEEEauthorblockN{Marcus Windmark}
  \IEEEauthorblockA{
  	mawi2661@student.uu.se \\
	Department of Information Technology\\
	Uppsala University
  }
}


%


\maketitle


\begin{abstract}
In this study an Artificial Neural Network was trained to classify musical instruments, using audio samples transformed to the frequency domain. Different features of the sound, in both time and frequency domain, were analyzed and compared in relation to how much information that could be derived from that limited data. The study concluded that in comparison with the base experiment, that had an accuracy of 93.5\%, using the attack only resulted in 80.2\% and the initial 100 Hz in 64.2\%.



\end{abstract}


\IEEEpeerreviewmaketitle

\section{Introduction}
Musical instruments come in a wide spectrum of shapes and sizes and the characteristics of one's sound can just as well be distinct or similar to another instrument~\cite{instruments}. 

The materials it was built with, the aging of its material, how the material was processed and even the style of the musician playing it may have an impact on how it sounds. An amateur comparing two instruments, built with different materials, may hear a difference between them, yet is still able to determine that they are the same type of instrument.

In this study, machine learning techniques were used to compare different characteristics of musical instruments and study their ability to distinguish a range of instruments. This was performed using the frequency spectrum of the audio signal, together with an Artificial Neural Network~\cite{artificial-neural-network} (ANN), comparing the accuracy of the network in the cases of the following experiments:

\begin{itemize}
\itemsep0.2em 
\item The whole sample
\item The attack of the sound
\item Everything but the attack of the sound
\item The initial 100 Hz of the frequency spectrum
\item The following 900 Hz of the frequency spectrum
\end{itemize}

\section{Feature Selection}
\label{sec:feature-selection}

The process of feature selection is an essential part to be able to use all information contained in the data. This section outlines the different possible use cases of the frequency spectrum that can be used as feature vectors.


\subsection{Base Experiment}
The feature vector that was used for the base experiment was constructed using the frequency spectrum of the original audio sample. This spectrum was then represented as partitions, described further in~\cref{sec:frequency-chunks} about pre-processing the data, and the overall properties of it  was used to classify the instruments.

\begin{figure}[ht]
  \centering
  \includegraphics[scale=0.25]{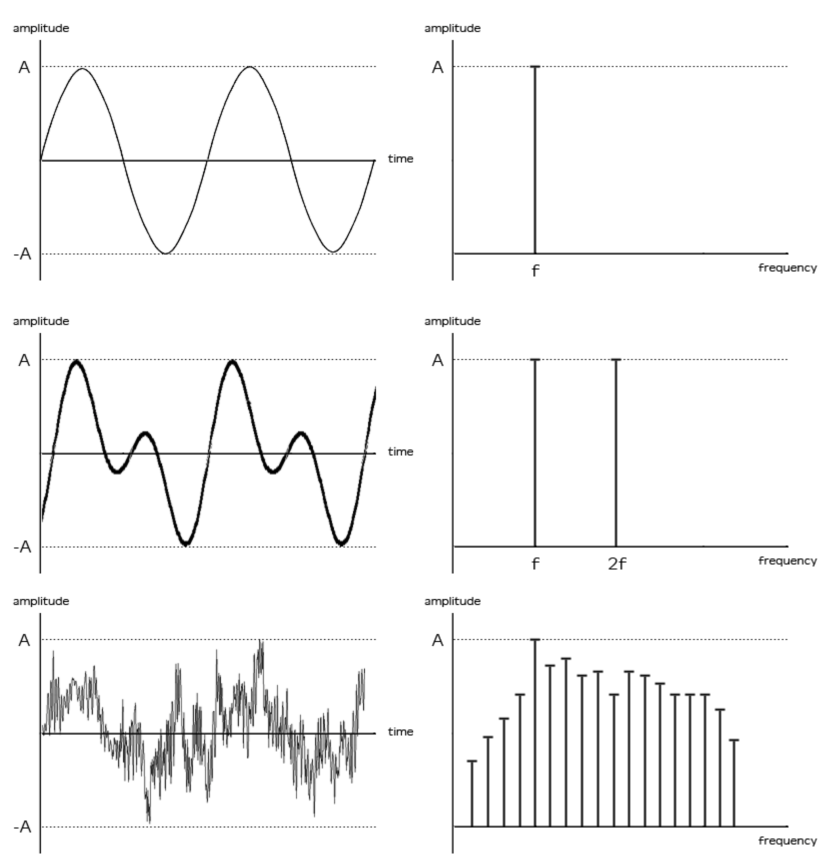}
  \caption{A comparison of representing different signals in time and frequency domain.}
  \label{fig:signal-time-frequency}
\end{figure}

The reason for using the representation of frequency domain as the basis for the feature vector, rather than the time domain, was because the frequency spectrum is a better way to represent and compare audio signals in, without taking length into consideration. Seen in figure~\ref{fig:signal-time-frequency} is an example of how different signals differ in the frequency spectrum of the signal. The frequency spectrum can be used as a discrete way of representing something that in reality is continuous, simplifying the process of computing calculations and doing pre-processing~\cite{broughton2011discrete}.

\subsection{Instrument Attack}
\label{sec:feature-attack}

\begin{figure}[ht]
  \centering
  \includegraphics[scale=\figureWidth]{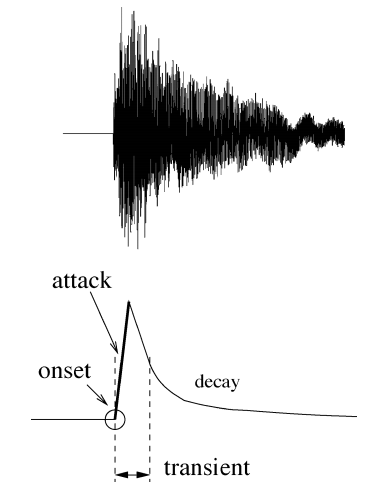}
  \caption{The beginning of the sound of an instrument consists of an onset, attack, transient period and decay.\protect\cite{bello05}}
  \label{fig:attack-explained}
\end{figure}

Many features characterize instruments and one important aspect is the attack transient. 
As shown in figure~\ref{fig:attack-explained}, the transient of an instrument consists of several parts. Starting from the onset point, the attack is the significant rise of the sound, ending in a longer decay period.

Clark~\cite{clark63} performed a study on the importance of the different parts of a tone for human recognition and concluded that having only the attack resulted in a good accuracy of recognizing most instruments. The hypothesis of this study was that the same would apply for an automated system and analyzing the importance of the attack by having only the attack as well as not having it at all.

\subsection{Low and High Frequencies}
\label{sec:lowhighfreq}
Following the standard of the Equal-tempered scale as described by Michigan Technology University\cite{physicsofmusic}, all of the audio samples used to train the network were in the fourth octave, ranging the frequency spectrum of the tones between 261.63 Hz ($C_{4}$) and 493.88 Hz ($B_{4}$) (see \cref{tab:tones}). In order to study a larger spectrum than just the tone frequency, and at the same time limit the scope of the experiment, only the frequencies ranging between 1 and 1000 Hz were used to construct the feature vector.

\begin{table}[ht]
\centering
\begin{tabular}{lc}
\multicolumn{1}{l}{\textbf{Tone}} & \multicolumn{1}{l}{\textbf{Frequency (Hz)}}  \\
$C_{4}$ 				&			261.63		\\
$C^{\#}_{4}$		& 			277.18 		\\
$D_{4}$				&			293.66			\\
$D^{\#}_{4}$				&			311.13		\\
$E_{4}$				&			329.63	 		 	\\
$F_{4}$				&			349.23			 	\\
$F^{\#}_{4}$				&			369.99			\\
$G_{4}$				&			392.00			\\
$G^{\#}_{4}$				&			415.30			\\
$A_{4}$				&			440.00			\\
$A^{\#}_{4}$				&			466.16			\\
$B_{4}$				&			493.88			\\  \\
\end{tabular}
\caption{Frequency range of the tones in the samples used to train the network.}
\label{tab:tones}
\end{table}

\section{Dataset}

\begin{figure}[ht]
  \centering
  \includegraphics[scale=0.46]{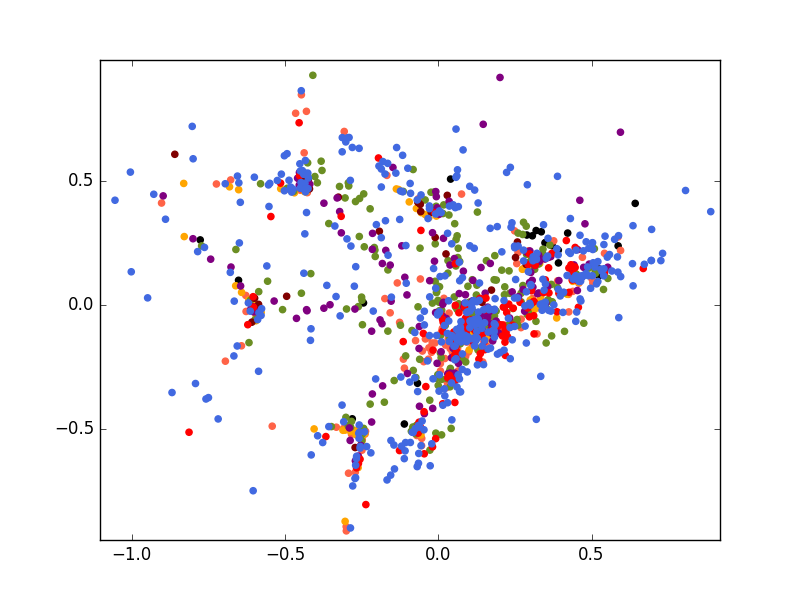}
  \caption{Principal Component Analysis plot - showing Banjo in black, Cello in pale red, Clarinet in orange, English horn in green, Guitar in maroon, Oboe in purple, Trumpet in red and Violin in royal blue.}
  \label{fig:dataset-scatter}
\end{figure}

The dataset used in this report was the London Philharmonic Orchestra dataset~\cite{dataset}, consisting of recorded samples from 20 different musical instruments. For each instrument, the samples range over its entire set of tones played in every octave with different levels of strength (\textit{piano}, \textit{forte}) and length. In addition to that, the dataset also includes samples where different playing techniques are used with the instrument, such as \textit{vibrato}, \textit{tremolo}, \textit{pizzicato} and \textit{ponticello}.

In order to limit the scope of this project, the following eight instruments were selected to train the model: Banjo, Cello, Clarinet, English horn, Guitar, Oboe, Trumpet and Violin. This set of instruments was chosen because of the high quality of the samples and them ranging over the three instrument families Brass, String and Woodwind. The number of samples of each instrument is shown in \cref{tab:dataset-samples}.

To avoid handling potential different harmonics in the same tone across the octaves, only the  samples of recordings done in the fourth octave were used.

\begin{table}[ht]
\centering
\begin{tabular}{llc}
\multicolumn{1}{l}{\textbf{Index}} & \multicolumn{1}{l}{\textbf{Instrument}} & \multicolumn{1}{c}{\textbf{Samples}} \\
1 				&			Banjo			& 			23  	\\
2 				& 			Cello			& 			166  	\\
3 				&			Clarinet		& 			131  	\\
4 				&			English horn	& 			234  	\\
5				&			Guitar	 		& 			29  	\\
6				&			Oboe			& 			155  	\\
7				&			Trumpet			& 			140  	\\
8				&			Violin			& 			366  	\\
\textbf{Total}	& & 	\textbf{1244 samples}  	\\ \\
\end{tabular}
\caption{Distribution of the instrument samples in the dataset.}
\label{tab:dataset-samples}
\end{table}

\subsection{Pre-processing of Base Experiment}
\label{sec:preprocess-base}

\begin{figure}[ht]
  \centering
  \includegraphics[scale=\figureWidth]{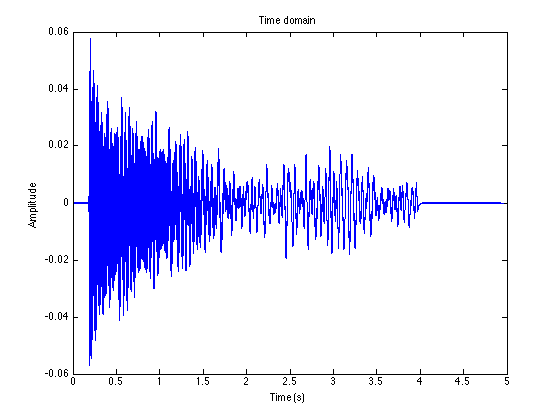}
  \caption{Unprocessed sample in time domain of a guitar playing the tone E4.}
  \label{fig:unprocessed-sample}
\end{figure}

Before training the network, the audio samples were pre-processed, as discussed in \cref{sec:feature-selection}. Initially, since the audio samples in time domain are continuous (as shown in \cref{fig:unprocessed-sample}), they had to be transformed into a discrete representation.

\subsubsection{Fast Fourier Transform}
\begin{figure}[ht]
  \centering
  \includegraphics[scale=\figureWidth]{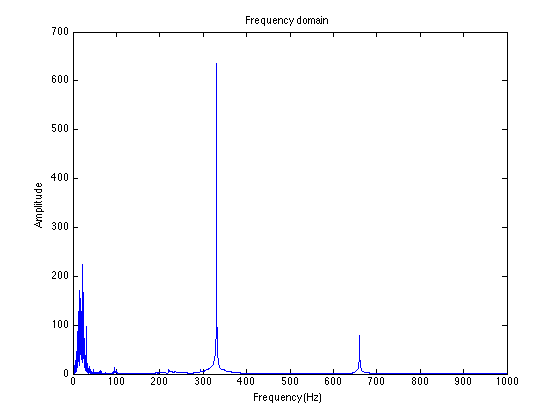}
  \caption{The sample in figure~\ref{fig:unprocessed-sample} transformed to the frequency domain.}
  \label{fig:fft-sample}
\end{figure}

The first step of the pre-processing consisted of transforming the audio sample from time domain to frequency domain by using the Fast Fourier Transform~\cite{weisstein2015fast} (FFT), resulting in a frequency spectrum  (see example in~\cref{fig:fft-sample}). The spectral components played an essential part in the further steps creating the feature vector.

\subsubsection{Spectrum Cut}
As motivated in \cref{sec:lowhighfreq}, the resulting frequency spectrum was cut off for all frequencies above 1000 Hz, leaving the range of 1-1000 Hz.

\subsubsection{Frequency shift}
\begin{figure}[ht]
  \centering
  \includegraphics[scale=\figureWidth]{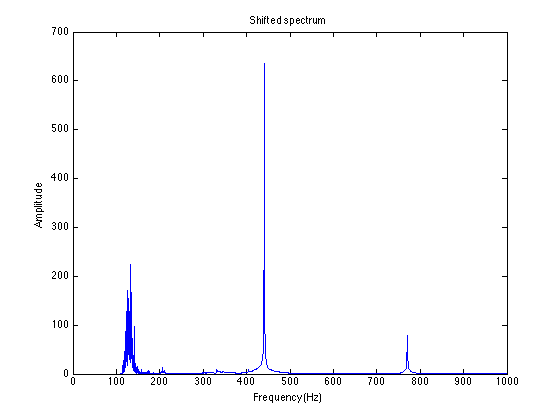}
  \caption{The audio sample with a shifted spectrum.}
  \label{fig:shift-sample}
\end{figure}

Since the purpose of network was not to learn, or even take into consideration, that each different tone has different base frequencies, all transformed samples were also shifted to be represented in the pitch of an $A_{4}$ (440 Hz). An example is shown in \cref{fig:shift-sample}. This resulted in audio samples regardless of tone having a similar representation and therefore minimizing the risk of the network learning to classify different tones rather than classifying musical instruments.

\subsubsection{Partitioning the Frequency Spectrum}
\label{sec:frequency-chunks}
\begin{figure}[ht]
  \centering
  \includegraphics[scale=\figureWidth]{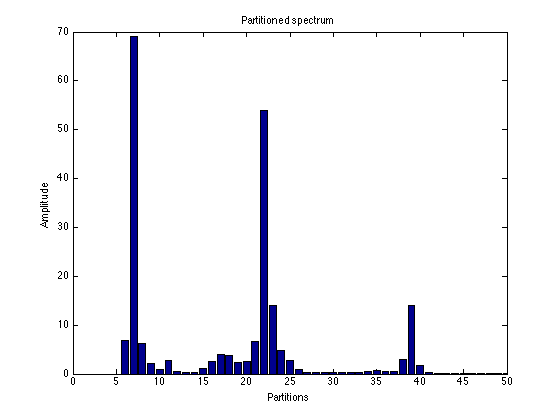}
  \caption{The partitioned frequency spectrum, where each section represented the average amplitude of that frequency range.}
  \label{fig:partition-sample}
\end{figure}

Having the frequency spectrum as basis, it was partitioned into ranges of frequencies, avoiding the inefficient case that would have been one feature per hertz. The process of partitioning was also to avoid a potential risk of overfitting the model. The spectrum was divided into 50 sections and each section was represented by the average frequency of that range, as seen in \cref{fig:partition-sample}.

\subsubsection{Normalization}
\begin{figure}[ht]
  \centering
  \includegraphics[scale=\figureWidth]{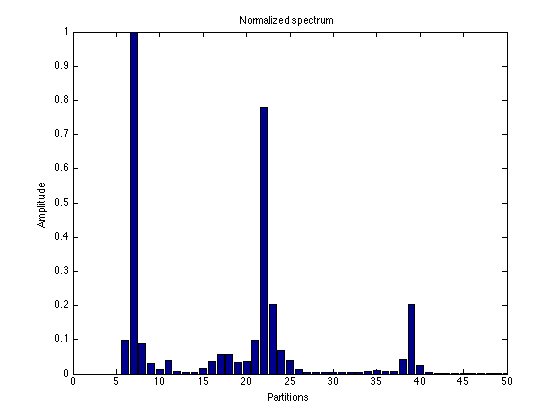}
  \caption{Frequency domain}
  \label{fig:sample-norm}
\end{figure}

In order to train a network on the co-relation between the amplitude tops of a sample and not the actual values, the data was normalized according to the feature scaling method, scaling the range of amplitudes between [0,1] (see \cref{eq:norm}). The normalization resulted in an emphasized co-relation of a sample's amplitudes across the frequency spectrum, allowing each sample to equally contribute to the training of the model (see \cref{fig:sample-norm}). 

\begin{equation}
x'  =  \frac{x-\text{min}(x)}{\text{max}(x) - \text{min}(x)}
\label{eq:norm}
\end{equation}

\subsection{Pre-processing of Comparative Experiments}
Extending the pre-processing of the base experiment described in \cref{sec:preprocess-base}, the pre-processing of the comparative datasets slightly differed in order to be used in the different experiments.

\subsubsection{Attack experiment}
To produce the feature vectors used in the experiments focusing on the attack, a specific technique of attack extraction was performed. This process was made in the time domain, before the transformation using FFT.

To find the onset point of the audio signal, it was divided into windows, each consisting of 10 ms of data. For each window, the Root Mean Square (RMS) energy was calculated and summed up. Using these discrete energy partitions, each was compared to the RMS energy of the overall signal and the onset was defined as the point where this energy was 10 dB over the signal average. 

There are many algorithms for finding the true length of the transient period (or attack), but considering that the diverse set of instruments in this project would have resulted in finding the onset to be a project on its own, the best alternative was to have a fixed transient length. Having a transient length of around 80 ms was shown in a study by Iverson~\cite{iverson93} to capture the important aspects of the onset in experiments with human subjects. In this project though, using a little longer transient length proved to work better, so the fixed transient length used was 100 ms.

In the experiment without an attack, the start of the steady period was shifted to be 200 ms after the end of the attack. This was done to be sure to totally exclude all characteristics from the attack.

\begin{figure}[ht]
\centering
  \begin{tabular}{cc}
	\includegraphics[scale=0.20]{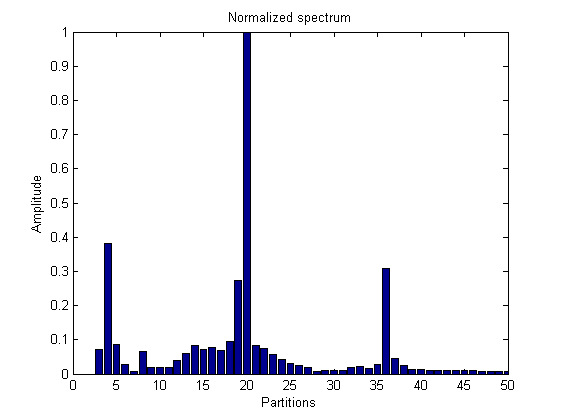} & 
 	\includegraphics[scale=0.20]{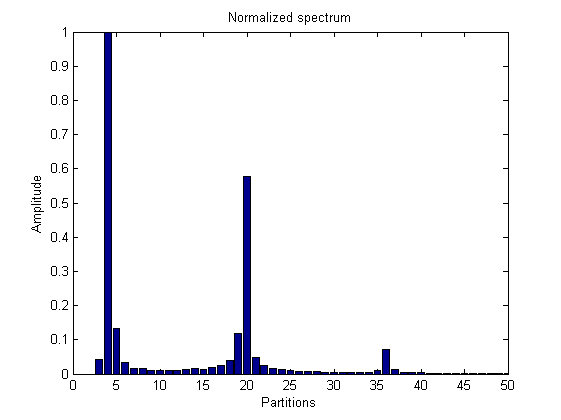} \\
 	\textit{Only attack} & 
  	\textit{Without attack}
  \end{tabular} 
  \caption{Comparison plots of feature vectors with or without attack.}
  \label{fig:compattack}
\end{figure}

A comparison of the frequency spectrum of both the attack and without attack can be seen in \cref{fig:compattack}.

\subsubsection{Frequency spectrum experiment}
The difference between the pre-processing of the base experiment and the frequency spectrum experiment was that the amplitudes outside of the range considered in the experiments were zeroed. In the case of the experiment containing the first 100 Hz, only this range of the data was preserved, similarly for the experiment with the following 900 Hz.

\begin{figure}[ht]
\centering
\begin{tabular}{cc}
	\includegraphics[scale=0.20]{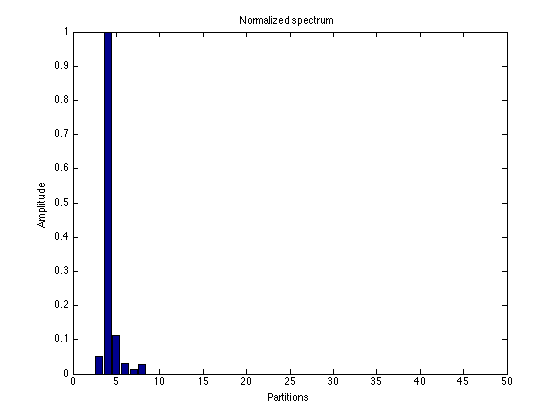} & 
 	\includegraphics[scale=0.20]{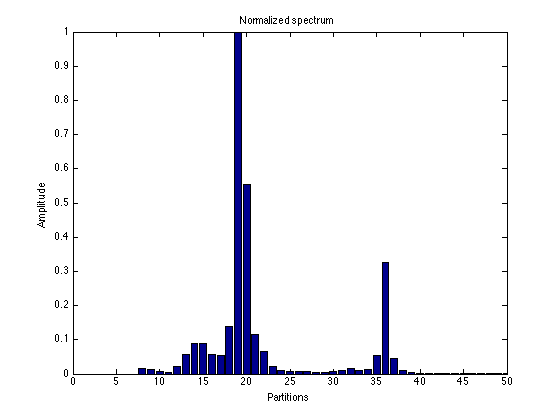} \\
 	\textit{The first 100 Hz} & 
  	\textit{The following 900 Hz}
  \end{tabular} 
  \caption{Comparison plots of feature vectors with initial 100 Hz and the following 900 Hz.}
   \label{fig:compinitialhz}
\end{figure}

\section{Training}
The model used for training was a Multilayer Perceptron~\cite{neural-networks-foundation} (MLP) with Early Stopping~\cite{prechelt1998automatic}, using Resilient Back Propagation~\cite{riedmiller1994rprop} (Rprop) as the learning heuristic. The network consisted of 50 inputs, 30 hidden nodes and eight different outputs, each representing one of the instruments.

\subsection{Learning Algorithm}
Resilient Back Propagation is a learning heuristic used for supervised learning. It takes only the sign of the partial derivative over all patterns into account (see \cref{eq:rprop}), and then acts independently on each weight update, shown in \cref{eq:rpropw}.

\begin{equation}
w_{ij}^{(t+1)}  =  w_{ij}^{(t)} + \Delta w_{ij}^{(t)} 
\label{eq:rprop}
\end{equation}

\begin{equation}
\Delta w_{ij}^{(t)} =  \left\{\begin{matrix}
 -\Delta _{ij}^{(t)}  & ,  & \text{if } {\frac{\partial E}{\partial w_{ij}}}^{(t)}  > 0 \\
+\Delta _{ij}^{(t)}  & ,  & \text{if } {\frac{\partial E}{\partial w_{ij}}}^{(t)}  < 0 \\
 0 &, & \text{else}
\end{matrix}\right.
\label{eq:rpropw}
\end{equation}

\subsection{Multilayer Perceptron}
\begin{figure}[ht]
\centering
\begin{tikzpicture}[shorten >=1pt,->,draw=black!50, node distance=\layersep]
    \tikzstyle{every pin edge}=[<-,shorten <=1pt]
    \tikzstyle{neuron}=[circle,fill=black!25,minimum size=17pt,inner sep=0pt]
    \tikzstyle{input neuron}=[neuron, fill=gray!40];
    \tikzstyle{output neuron}=[neuron, fill=gray!40];
    \tikzstyle{hidden neuron}=[neuron, fill=black!60];
    \tikzstyle{annot} = [text width=4em, text centered]

    \foreach \name / \y in {1,...,2}
        \node[input neuron, pin=left:\footnotesize{Input \#\y}] (I-\name) at (0,-\y) {};

    \foreach \name / \y in {1,...,3}
        \path[yshift=0.5cm]
            node[hidden neuron] (H-\name) at (\layersep,-\y cm) {};

    \node[output neuron,pin={[pin edge={->}]right:\footnotesize{Output}}, right of=H-2] (O) {};

    \foreach \source in {1,...,2}
        \foreach \dest in {1,...,3}
            \path (I-\source) edge (H-\dest);

    \foreach \source in {1,...,3}
        \path (H-\source) edge (O);

    \node[annot,above of=H-1, node distance=1cm] (hl) {\scriptsize{Hidden layer}};
    \node[annot,left of=hl] {\scriptsize{Input layer}};
    \node[annot,right of=hl] {\scriptsize{Output layer}};
\end{tikzpicture}
    \caption{Basic MLP setup}
\label{fig:mlp}
\end{figure}
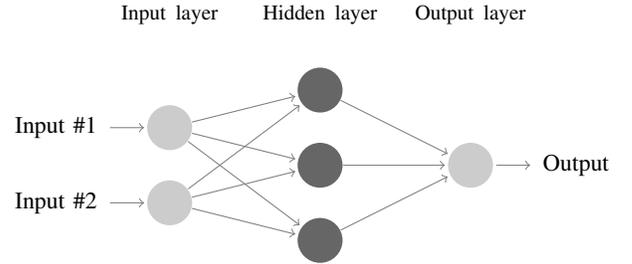

Multilayer Perceptrons~\cite{neural-networks-foundation} (MLP) are feed forward ANNs where sets of inputs are mapped to appropriate outputs using layers of nodes in a directed graph (see \cref{fig:mlp}). The nodes in the layers consist of perceptrons, binary classifiers with a function deciding whether an input belongs to one class or the other using \cref{eq:percone} to \cref{eq:percfour}.

\begin{equation}
y=f_{h}(S) 
\label{eq:percone}
\end{equation}

\begin{equation}
f_{h}(S) = \left\{\begin{matrix}
1, S > 0\\ 
1, S \leq 0
\end{matrix}\right. 
\label{eq:perctwo}
\end{equation}

\begin{equation}
  S = \sum_{n}^{i=1}w_{i}x_{i} - \theta = 
\label{eq:percthree}
\end{equation}

\begin{equation}
= \sum_{n}^{i=0}w_{i}x_{i} \ \ \ \text{where} \ \ \ \left\{\begin{matrix}
x_{0} = -1\\ 
w_{0} = \theta
\end{matrix}\right.
\label{eq:percfour}
\end{equation}

\subsection{Hidden Nodes}
Using 30 hidden nodes in a network could in some cases lead to inefficiency and overfitting, but due to the fact that the amplitude in the frequency spectrum had large variances and that the data was normalized, many of the elements would be close to 0. Therefore, the risk of overfitting is greatly reduced, as many of the inputs would most likely not have a large effect on the error, and thus necessarily not negatively influence the weight changes in the training.

\begin{table}[ht]
\centering
\begin{tabular}{lc}
\multicolumn{1}{l}{\textbf{Data split}} & \multicolumn{1}{c}{\textbf{Size}} \\
Training data		& 			60\% 	\\
Validation data		& 		20\% 	\\
Test data		& 				20\% 	\\  \\
\end{tabular}
\caption{How the data was distributed in Early Stopping}
\label{tab:split-percentages}
\end{table} 

\subsection{Early Stopping}
Early Stopping~\cite{prechelt1998automatic} is a regularization method to avoid overfitting by splitting the data into three subsets: training, validation and test. The network uses the training set to improve its performance while validating with the validation set, up to the point that the validation error is increasing. It then retrains the network with the training data up to the epoch where the validation error was at a minimum, and then uses the test set to test its performance.

To optimize the training accuracy and to avoid overfitting, the network was trained using Early Stopping~\cite{prechelt1998automatic}. The maximum fail parameter was set to 150 epochs and the total numbers of epochs were set to 500. The dataset was then split as shown in \cref{tab:split-percentages}.

\section{Experimental Results}
\label{sec:results}
As stated in \cref{sec:feature-selection} about the feature selection, five different experiments were performed. A summarized result of the accuracy is shown in \cref{tab:result-accuracy}. In this section, we are going to present the results from the experiments and then have a further discussion in \cref{sec:discussion}. 

The accuracy of each experiment was measured using an average of 10 runs of each experiment.

\begin{table}[ht]
\centering
\begin{tabular}{lc}
\textbf{Experiment} & \textbf{Accuracy}\\
Base experiment	&		93.5\%  	\\
Only attack		&		80.2\%  	\\
Without attack	&		73.2\%  	\\
First 100Hz		& 		64.2\%  	\\
Following 900Hz		&		90.6\%  	
\\ \\
\end{tabular}
\caption{The average of six times result of the five different experiments ranged from 66\% to 94\%.}
\label{tab:result-accuracy}
\end{table}

\subsection{Base Experiment}
\begin{figure}[ht]
  \centering
  \includegraphics[scale=\confusionMatrixWidth]{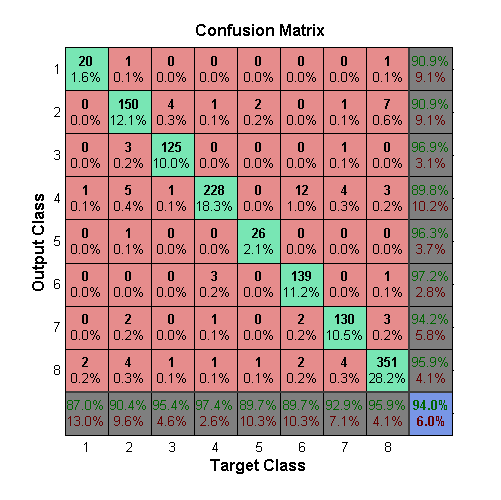}
  \caption{The confusion matrix from one of the training sessions of the base experiment.}
  \label{fig:confusion-base-experiment}
\end{figure}

\begin{figure}[ht]
  \centering
  \includegraphics[scale=0.61]{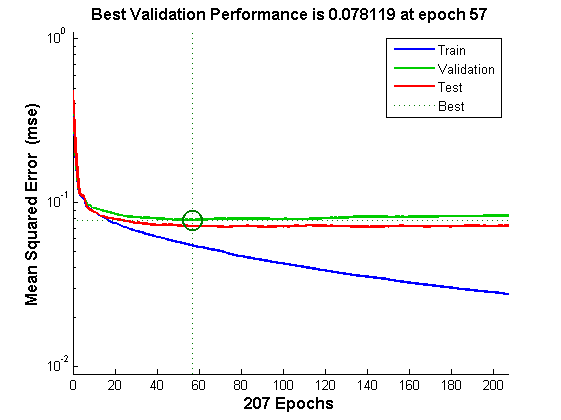}
  \caption{Early Stopping plot from one of the training sessions of the base experiment.}
    \label{fig:early-stopping-base-experiment}
\end{figure}

Training the network with the base feature vector resulted in an average accuracy of 93.5\% and a sample confusion is shown in \cref{fig:confusion-base-experiment}. An example of the error performance of the early stopping algorithm is shown in \cref{fig:early-stopping-base-experiment}, stopping the training of the network after 207 epochs.

\subsection{Impact of the Attack}
\begin{figure}[ht]
  \centering
  \includegraphics[scale=\confusionMatrixWidth]{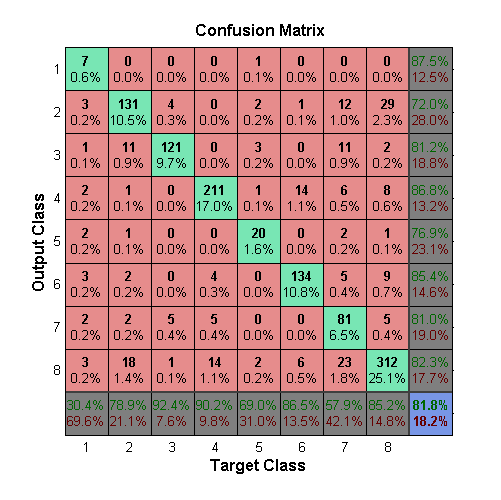}
  \caption{The confusion matrix from one of the training sessions using only the attack to train the network.}
  \label{fig:confusion-only-attack}
\end{figure}

\begin{figure}[ht]
  \centering
  \includegraphics[scale=\confusionMatrixWidth]{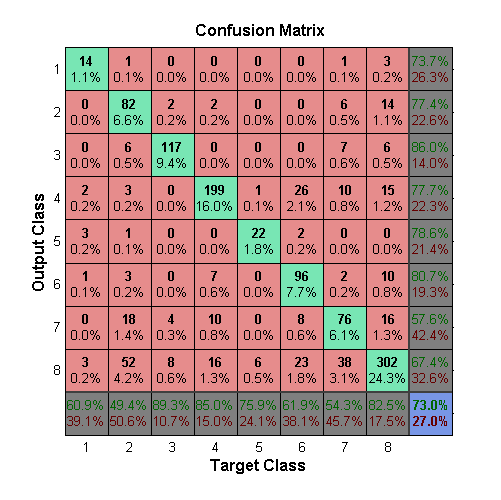}
  \caption{The confusion matrix from one of the training sessions excluding attack to train the network.}
  \label{fig:confusion-without-attack}
\end{figure}

The network was trained with the two different aspects of feature vectors either containing only the attack or excluding it. The feature vector using only the attack resulted in an accuracy of 80.2\%, as seen in \cref{fig:confusion-only-attack}, and the result of excluding the attack had an accuracy of 73.2\%, shown with an example in \cref{fig:confusion-without-attack}.

\subsection{Impact of the Frequency Cut}
The network was trained with two different versions of feature vectors, one with the initial 100 Hz and another with the following 900 Hz. 

\begin{figure}[ht]
  \centering
  \includegraphics[scale=\confusionMatrixWidth]{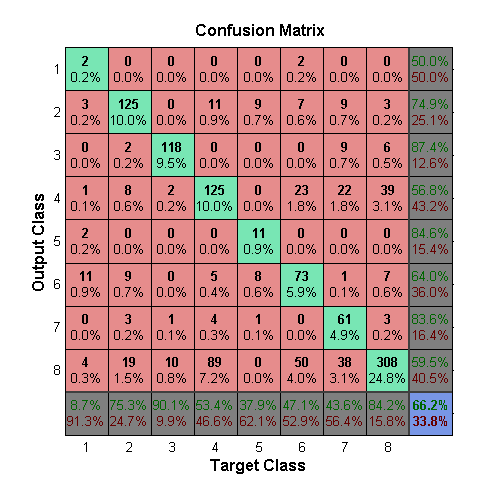}
  \caption{The confusion matrix from one of the training sessions using the initial 100 Hz to train the network.}
  \label{fig:cfonehundred}
\end{figure}

\begin{figure}[ht]
  \centering
  \includegraphics[scale=\confusionMatrixWidth]{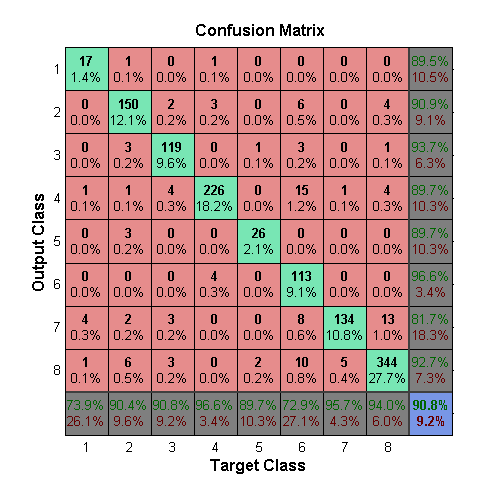}
  \caption{The confusion matrix from one of the training sessions using following using 900 Hz to train the network.}
  \label{fig:cfninehundred}
\end{figure}

Training the network with the initial 100 Hz resulted in an accuracy of 64.2\% (\cref{fig:cfonehundred}),
in comparison to training the network with the following 900 Hz, which resulted in an accuracy of 90.6\% (see \cref{fig:cfninehundred}).

\section{Discussion}
\label{sec:discussion}

The result of the base experiment, as presented in the confusion matrix in \cref{fig:confusion-base-experiment}, had an overall accuracy of 93.5\%. Considering that this was a simple experiment based only on partitioning and normalizing the frequency spectrum, the high accuracy was surprising. 

\begin{figure}[ht]
  \centering
  \includegraphics[scale=0.4]{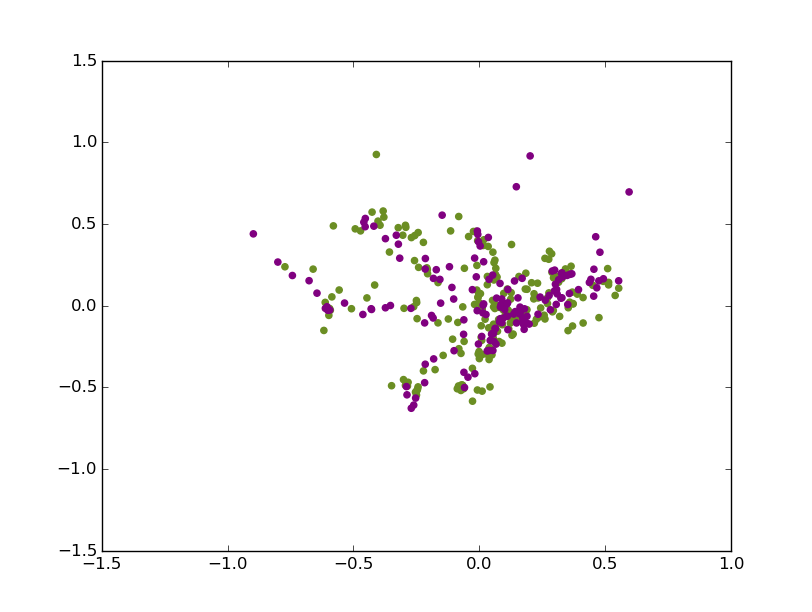}
  \caption{Scatter plot of the Principal Component Analysis limited to only English horn in green and Oboe in purple shows the features almost totally overlapping.}
  \label{fig:oboe-vs-enghorn-scatter}
\end{figure}

Inspecting the confusion matrix in \cref{fig:confusion-base-experiment}, the most difficult instrument to predict was English horn, being incorrectly predicted as an Oboe. This may not come as a surprise, since these who instruments are both of the woodwind family and hard to distinguish even for most non-musicians. Shown in \cref{fig:oboe-vs-enghorn-scatter} is the scatter plot of the dataset, as shown in \cref{fig:dataset-scatter}, but limited to only these two instruments. The features have a major overlap, with no definite distinction between the two classes. This shows that the classifying network had a hard time utilizing the spectrum information for the feature vector defined in this experiment.

\subsection{The Impact of the Attack}

Comparing the two experiments using either only the attack or totally excluding it as feature vectors confirmed the hypothesis of the importance of the attack, as discussed in \cref{sec:feature-attack}. Inspecting the results, the difference in accuracy was 80.2\% compared to only 73.2\%. 

As seen in the confusion matrices of the experiments in \cref{fig:confusion-only-attack} and \cref{fig:confusion-without-attack}, there are several instruments with a clear classification advantage for the attack feature vector. 

\begin{figure}[ht]
\centering
  \begin{tabular}{cc}
	\includegraphics[scale=0.27]{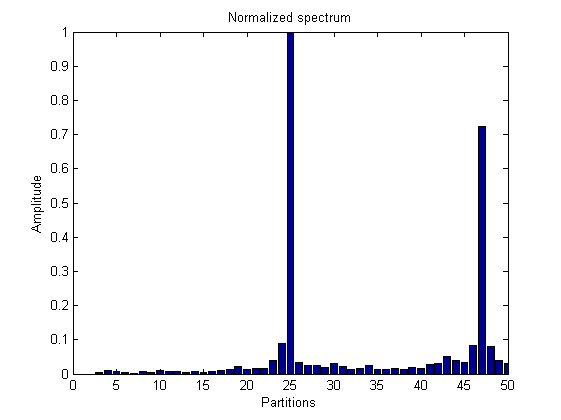} & 
 	\includegraphics[scale=0.27]{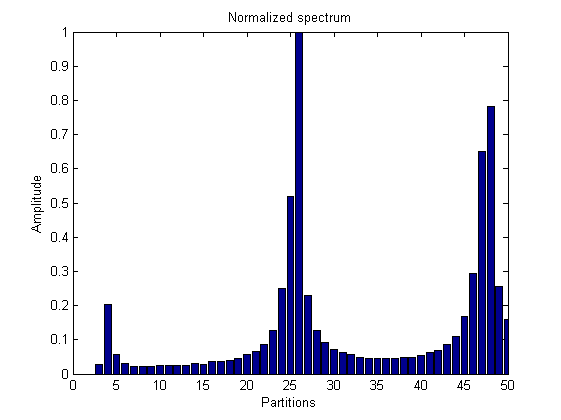} \\
 	\textit{Only attack} & 
  	\textit{Without attack}
  \end{tabular} 
  \caption{Comparison plots of feature vectors with or without attack in a sample file of trumpet.}
  \label{fig:compattack-trumpet}
\end{figure}

The biggest difference was the classification of trumpet, having an accuracy of 81\% with the attack and only 57.6\% without. Comparing the spectrum between the two experiments, the frequency amplitudes in each case have distinct features, as shown in \cref{fig:compattack-trumpet}. Besides the two clear frequency tops present in both experiments, having no attack resulted in many more frequencies with lower amplitudes in between these tops. One reason for the low accuracy achieved in this case could be due to all this noise, resulting in less distinction of the important frequencies.

\subsection{The Impact of the Initial Frequency Spectrum}
One conclusion the accuracy of training the network with only the initial 100 Hz of the spectrum says is that it seems to hold valuable information. It is represented in a feature vector with a magnitude of a five times less information than the feature vector used in the base experiment, but drops only 29.3\% percentage points (to a 64.2\% accuracy). 

Whether 63.2\% accuracy is good can be considered debatable, but there is still some information held in to the lower frequencies that opens up possibilities to further study in future work (see \cref{sec:futurework-initial}).

\subsection{Possible Introduction of Bias}
\label{sec:disc-bias}
In this experiment, since raw data was both modified and limited in order to construct the feature vector, there is a risk that some bias was introduced that could have impacted the results.

For example, only using the initial 1000 Hz of the frequency spectrum may cause loss of information regarding overtones or harmonics that potentially could improve the results in either of the experiments.

Also, since the frequency spectrum is normalized, all information regarding the actual amplitudes is lost in trade for how the amplitudes of the frequencies for a single sample co-relates. There may be some information in the amplitudes that would further improve the results of either of the experiments.

\section{Future Work}
The results of the performed experiments bring us some  conclusions that could be further studied.

\subsection{Further Studying the Attack}
The current experiments using the attack are all very simple in their nature, by only separating the attack and then creating the normalized feature vector of the frequency spectrum. An interesting future work would be to use the characteristics of the attack itself and from that create specialized features. The shape of the attack is an important factor and using features such as rise and decay time, as well as the derivative of them both, would possibly help classify instruments. 

Other features of the attack include the RMS energy, as well as the rise time from the point of onset to the point where the RMS energy is maximized. Having the feature vector contain several of these characteristics, or alternatively using them together with the currently used normalized spectrum, would lead to many interesting experiments and possible extensions.

\subsection{Further studying the initial 100 Hz}
\label{sec:futurework-initial}
As discussed in~\cref{sec:discussion}, the initial 100 Hz seems to hold some key information that can be used in musical instrument recognition. Though it did not outperform the following 900 Hz, one could argue that there is still an interesting aspect of how well it performed considering that its accuracy dropped 29.3\% percentage points when the information in the feature vector 90\% smaller. Perhaps a different training method could be used, or the data pre-processed differently that would impact the accuracy.

\subsection{Expanding the Studied Frequency Range}
As pointed out in~\cref{sec:disc-bias}, there may be some information of overtones or harmonies beyond 1000 Hz that was overlooked in this experiment.  Further studies on this experiment could potentially explore if expanding the range of frequencies used to construct the feature vector possibly can improve the accuracy of the network. In some related work (see \cref{sec:relatedwork}) show examples of using up to 1500 Hz to train their models.

\subsection{Using the Mel-frequency Cepstrum}

Mel-frequency cepstrum coefficients (MFCC) are commonly used in the application of speech recognition and are also applicable in classifying musical instruments. The essential part of MFCC is taking the combining the discrete cosine transform with the power spectrum of the Fourier Transform of a signal~\cite{brown99}. The MFCC has been used by some studies to test the effectiveness of using the cepstrum to classify instruments; so applying this alternative view of the spectrum to the experiments in this study would lead to an interesting comparison.

\section{Related work}
\label{sec:relatedwork}

\subsection{Human Recognition of Audio Signals}

There have been a number of studies on the human ability to distinguish musical instruments from each other by investigating the auditory properties of the sound itself. McAdams and Bigand~\cite{mcadams1993thinking} identified three parts of the musical sound event (attack, middle sustain and final decay) and compared how they did, or did not, impact in humans identifying the instrument depending on external factors. For example, they studied the importance of the attack by noticing the large drop in performance when the attack was cut out, but also how this reduction minimized when vibrato was used. Therefore, drawing the conclusion that the most important information of an instrument exists in the first part of the sound event, but in absence of that  information, additional information still existed in the sustain portion is augmented slightly when changes to the pattern that specify the resonance structure of the source was present, such as vibrato.

\subsection{Musical Instrument Recognition Systems}

There have been many studies performed on musical instrument recognition. When classifying instruments, there are many characteristics defining them that can be used. The studies described in this section used a range of different features and they can all be divided into the two categories spectral (frequency) and temporal (time).

Early studies used statistical pattern-recognition techniques. An early example is the study performed by Bourne~\cite{bourne72} 1972, which used spectral data of the sound as input to a Bayesian classifier to distinguish between three instruments. The same methods have been applied later as well; Fujinaga~\cite{fujinaga98} used properties of the steady-state of the sound together with a k-nearest neighbor~\cite{cunningham2007k} (kNN) classifier, achieving an accuracy of 50\% using 23 instruments. 

In a study performed by Brown~\cite{brown01}, methods from the well-researched field of Speaker Identification were used to determine the most effective features to be able to separate four different instruments. Using features such as Cepstral coefficients, histogram differentials and cross correlation, an accuracy of around 80\% was obtained.

Eronen~\cite{eronen00} used a wide set of features, covering both spectral and temporal properties of the musical instruments, was used together with a Gaussian classifier in combination with kNN). Out of a total of 30 instruments, the instrument family was successfully classified with 94\% accuracy, whereas individual instruments had an accuracy of 80\%.

Kaminskyj~\cite{kaminskyj95} used the RMS energy envelope of the temporal spectrum as features to compare the performance of  ANNs with a simple kNN classifier, setting k to 1, to classify four instruments. Using kNN performed best, achieving a 100\% accuracy, compared to around 96\% in the case of ANN, which is motivated by kNN being able to take more info from the RMS into account than the 32 weights used in ANN.


\section{Conclusion}
The approach of this paper was to evaluate the features in a system performing musical instrument recognition. The five different experiments that were performed all used a feature vector of length 50, containing a normalized frequency spectrum of the audio signal.

Training an ANN to perform musical instrument recognition was highly successful,  and using a feature vector based on the first 1000 Hz of the frequency spectrum reached an average accuracy of 93.5\%. Also, studying  the impact of  certain smaller aspects in the sound, such as a more limited frequency range or a limited portion of the audio sample, seemed to prove that these aspects held  some key information that still enabled a fairly good distinction between different instruments with an accuracy at best being 80.2\%  and worst 64.2\%.

This study lays the foundation for interesting future work, further examining these aspects and experimenting with different representations of them to optimize accuracy.

\bibliographystyle{IEEEtran}

\bibliography{bibliography}

\end{document}